\documentclass{article}
\usepackage[utf8]{inputenc}
\usepackage{amsmath}
\usepackage{subcaption}
\usepackage{graphicx}
\usepackage{amsfonts}
\usepackage{amssymb}
\usepackage{algorithm}
\usepackage[noend]{algpseudocode}
\usepackage{times} % This package changes the default font to Times for better readability
\usepackage{arxiv}

\usepackage[utf8]{inputenc} % allow utf-8 input
\usepackage[T1]{fontenc}    % use 8-bit T1 fonts
\usepackage{hyperref}       % hyperlinks
\usepackage{url}            % simple URL typesetting
\usepackage{booktabs}       % professional-quality tables
\usepackage{amsfonts}       % blackboard math symbols
\usepackage{nicefrac}       % compact symbols for 1/2, etc.
\usepackage{microtype}      % microtypography
\usepackage{lipsum}         % Can be removed after putting your text content
\usepackage{graphicx}
\usepackage{doi}

\title{A Novel No-Reference Image Quality Metric for Assessing Sharpness in Satellite Imagery}

\author{Lucas Gonzalo Antonel \\
    Satellogic S.A. \\
    \texttt{lucas.antonel@satellogic.com} \\
}

\renewcommand{\headeright}{}
\renewcommand{\undertitle}{}
\renewcommand{\shorttitle}{}

\hypersetup{
pdftitle={A NOVEL NO-REFERENCE IMAGE QUALITY METRIC FOR
ASSESSING SHARPNESS IN SATELLITE IMAGERY},
pdfauthor={Lucas Gonzalo Antonel},
}
\date{} % This removes the date

\begin{document}
\maketitle

\begin{abstract}
This study introduces a novel no-reference image quality metric aimed at assessing image sharpness. Designed to be robust against variations in noise, exposure, contrast, and image content, it measures the normalized decay rate of gradients along pronounced edges, offering an objective method for sharpness evaluation without reference images. Primarily developed for satellite imagery to align with human visual perception of sharpness, this metric supports monitoring and quality characterization of satellite fleets. It demonstrates significant utility and superior performance in consistency with human perception across various image types and operational conditions. Unlike conventional metrics, this heuristic approach provides a way to score images from lower to higher sharpness, making it a reliable and versatile tool for enhancing quality assessment processes without the need for pristine or ground truth comparison. Additionally, this metric is computationally efficient compared to deep learning analysis, ensuring faster and more resource-effective sharpness evaluations.
\end{abstract}

% keywords can be removed

\section{Introduction}
The expansion in Earth observation satellite deployments has significantly increased the volume and diversity of collected imagery data, introducing challenges in maintaining consistent image quality. Manual visual inspection is impractical due to the vast quantity of data, necessitating automated methods to ensure high-quality imagery, particularly in terms of sharpness. Image sharpness can be affected by various factors such as vibrations, motion blur, and issues within the telescope system. This highlights the need for a rigorous and systematic methodology to assess image sharpness, ensuring the reliability and utility of satellite imagery across diverse applications.

To address this need, we propose a metric designed for on-orbit evaluation of satellite imagery sharpness. This metric is innovative in several key aspects: it requires no reference images, making it highly adaptable to the dynamic nature of Earth observation; it is directional, allowing for the assessment of sharpness across different image directions; and it is capable of evaluating images as part of a larger satellite constellation, providing a comprehensive overview of image quality across diverse operational contexts.

Moreover, this metric incorporates a statistical approach to account for the volume of images produced and the inherent variability in image content, ensuring broad applicability across different types of Earth observation missions. By automating the image quality assessment (IQA) process, our metric offers a scalable solution to the challenge of maintaining high-quality imagery in the face of increasing data volumes and complex operational environments. This approach not only facilitates effective monitoring of satellite imaging systems' performance but also provides a robust method for estimating sharpness as an emergent property, enabling more detailed analysis of the impact of subsystem improvements on overall image quality.
Additionally,  by normalizing the sharpness independently of the content, we were able to calculate the sharpness in different directions of the image. Specifically, in our case, we perform this analysis in the horizontal and vertical directions. This approach enables the diagnosis of the type of blur affecting the image and the satellite, such as motion blur or out-of-focus blur, among others.

\section{Overview of No-Reference Image Sharpness Assessment Techniques}

The sharpness evaluation methods for no-reference images are categorized into four main types: spatial domain-based methods, spectral domain-based methods, learning-based methods, and combination methods. Spatial domain methods utilize image characteristics in the spatial domain, such as grayscale gradients and edge detection, to differentiate between blurred and clear images \cite{Brenner1976, Marziliano2002, Zhou2018, Li2016}. Grayscale gradient functions, including the Brenner function and the Laplacian function, calculate differences between adjacent pixels to evaluate sharpness \cite{Zhan2020, Li2021}. Edge detection methods, using algorithms like the Canny and Sobel operators, focus on detecting edges but can be noise-sensitive \cite{Canny1986, Sobel1970, Prewitt1960, Marziliano2002, Zhang2016}. Spectral domain methods, which transform the image to the frequency domain via Fourier and wavelet transforms, can extract detailed edge information but often entail high computational complexity \cite{Kanjar2016, Bae2018, Baig2022, Kerouh2018, Vu2011, Hassen2013, Wang2018, Gvozden2018}. Fourier transform techniques include FFT, DFT, and DCT implementations \cite{Bae2018}.

Learning-based methods apply machine learning and deep learning to evaluate image sharpness by training on extensive datasets, achieving high accuracy and robustness \cite{Pei2015, Liu2016, Zhu2018, Lin2020}. Early approaches using support vector machines (SVM) and support vector regression (SVR) showed good performance on smaller datasets \cite{Moorthy2011, Pei2015}. Modern deep learning techniques, such as convolutional neural networks (CNN) and generative adversarial networks (GAN), have enhanced image quality prediction \cite{Zhu2018, Li2020, Lin2020, Gao2019}. Dictionary learning methods leverage sparse representation and clustering for sharpness evaluation \cite{Li2016, Xu2018, Jiang2019}. Combination methods, which integrate multiple techniques, demonstrate high accuracy and robustness \cite{Vu2011, Liu2018}. The rise of learning-based methods, driven by advancements in machine learning, is notable \cite{Zhu2023, Mittal2012}. Evaluations on public datasets using metrics like mean squared error (MSE) and structural similarity index (SSIM) highlight the effectiveness of learning-based and combination methods in correlating with human perception \cite{Moorthy2011, Bae2018}. Despite progress, challenges remain in developing universally applicable methods, with future research focusing on enhancing robustness, reducing computational complexity, and improving real-time performance \cite{Moorthy2011, Zhu2023}.

\section{Sharpness Metric Description}

We propose a method to evaluate image sharpness based on the assessment of grayscale gradients in the spatial domain. Our approach, aims for precision, low computational complexity, and directional sensitivity of the metric to diagnose potential causes of blur. The underlying intuition of the metric is to measure the difference between the most intense gradients of an image and those same gradients after applying a Gaussian blur. This yields a decay rate which, once normalized, indicates the percentage by which the gradients are affected by the blur. Our findings consistently show that the most intense gradients, representing the sharpest edges, exhibit a significantly higher decay rate if they are well-defined edges, whereas the decay rate is low for diffuse edges. Additionally, this approach with some adjustments demonstrates consistent results across different types of images, with varying brightness, content, contrast, and noise levels.

Equally important, the varying content of satellite images, combined with the limitations of the metric, led us to introduce a representative indicator. This indicator quantifies how useful an image is for computing sharpness, allowing us to filter out excessively cloudy images, those lacking edge content, or images with very high-frequency content (a case we will explain further later). In this way, we complemented our quality metric with a means of measuring the confidence in our sharpness assessment of the image using this method, thereby enhancing the precision of the final result. Selected images that surpass this threshold of representative indicator are then systematically used to monitor satellite performance, assess image quality, and guide decision-making regarding satellite operations. This methodological approach not only enhances the efficiency of satellite imagery analysis but also bases operational decisions on quantitative, data-driven assessments of image quality.

Now, we will show in detail the step-by-step process of the sharpness metric. The algorithm flow is as shown in the Figure 1.

\begin{figure}[htbp]
\centering
\includegraphics[width=\linewidth]{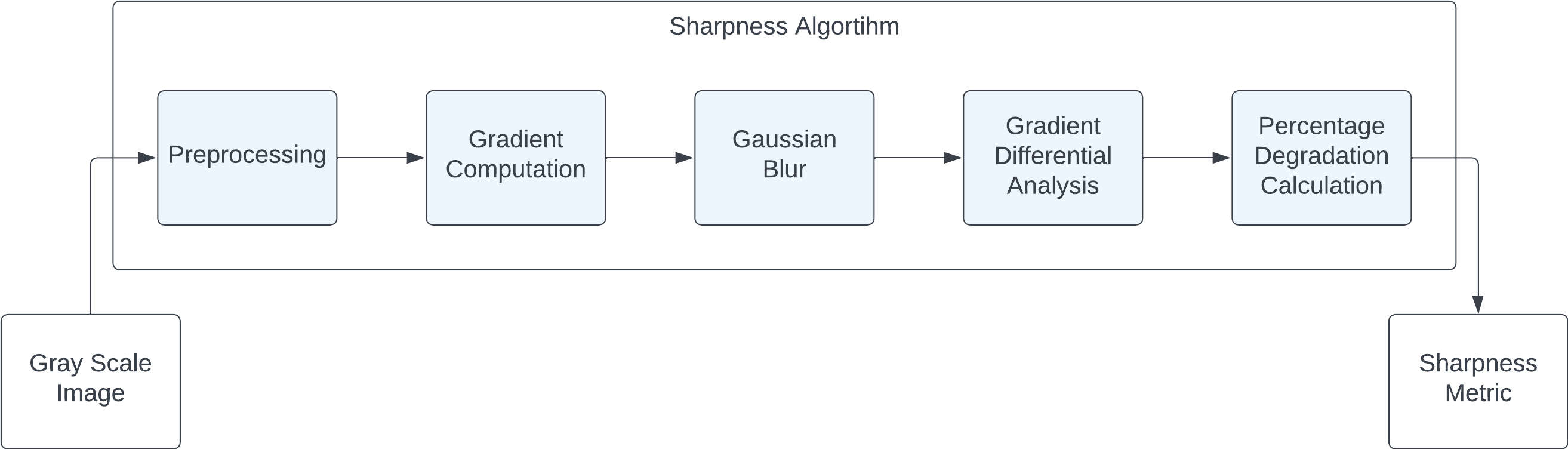}
\caption{Sharpness Algorithm Diagram} % Opcional
\label{fig:algo_diagram} 
\end{figure}

\subsection{Preprocessing}
The preprocessing filter out high frequency noise from the sensor. To achieve this, we target pixels exhibiting a specific percentage difference from their immediate neighbors and replace them with the average value of those neighbors. This process can be mathematically described as follows:

\begin{equation}
p'_{i,j} = 
\begin{cases} 
\frac{1}{N} \sum_{(k,l) \in \mathcal{N}(i,j)} p_{k,l} & \text{if } \left| \frac{p_{i,j} - \mu_{\mathcal{N}(i,j)}}{\mu_{\mathcal{N}(i,j)}} \right| > \theta, \\
p_{i,j} & \text{otherwise},
\end{cases}
\end{equation}

where \(p'_{i,j}\) denotes the value of the pixel after noise filtering, \(\mathcal{N}(i,j)\) represents the set of indices for the neighbors of pixel \((i,j)\), \(N\) is the number of pixels in \(\mathcal{N}(i,j)\) (excluding \((i,j)\) itself if deemed necessary), \(\mu_{\mathcal{N}(i,j)}\) is the average value of the pixels in \(\mathcal{N}(i,j)\), and \(\theta\) is the specified percentage difference threshold for identifying noise affected pixels. This approach effectively mitigates the impact of sensor noise by smoothing out anomalies in high frequency data.
The next step,  involves applying a mask \(M_{LH}\) to the image \(I\), which selectively filters pixels based on their intensity values, emphasizing those within a specified range and excluding the extreme values. Specifically, for an image with pixel values ranging from 0 to 255, \(M_{LH}\) filters out pixels with values at the extremes of this range, i.e., 0 and 255. This can be formally described by the following equation:
\begin{equation}
I_p = I \times M_{LH}(I)
\end{equation}
where \(I_p\) is the preprocessed image, \(I\) is the original image, and \(M_{LH}\) represents the low-high intensity mask that is applied to \(I\), designed to exclude pixel values at the boundary of the intensity range, enhancing the focus on pixels within the desired intensity interval. The objective of this stage is to filter out false edges resulting from saturation or areas with high noise content due to underexposure.

\subsection{Gradient Computation and Isolation}
After preprocessing, we calculate the gradients of the image, \(I_p\), to highlight its edge information. For this purpose, a \(5 \times 5\) Sobel filter is applied in both the horizontal and vertical directions, denoted as \(G_x\) and \(G_y\), respectively. The choice of the filter size was determined heuristically based on the image size and noise characteristics. The equations are given by:
\begin{equation}
G_x = \nabla_x(I_p) 
\end{equation}
\begin{equation}
G_y = \nabla_y(I_p)
\end{equation}
The gradients \(G_x\) and \(G_y\) represent the changes in intensity in the X and Y directions, respectively, accentuating edges by emphasizing the rate of change in intensity.

Subsequently, the gradients are filtered based on upper (\(P_U\)) and lower (\(P_L\)) percentile thresholds to isolate significant edges. This filtering process aims to highlight the most pronounced edges by focusing on gradient values within the 98.5th to 99.5th percentile range, determined heuristically in relation to the image size and noise. The filtered gradients are computed as follows:
\begin{equation}
G_{x,\text{filtered}} = G_x \times M_{Px}(G_x, P_{Ux}, P_{Lx})
\end{equation}
\begin{equation}
G_{y,\text{filtered}} = G_y \times M_{Py}(G_y, P_{Uy}, P_{Ly})
\end{equation}
where \(M_{Px}\) and \(M_{Py}\) are masks that retain gradients within the \(P_U\) and \(P_L\) percentile ranges for the X and Y directions, respectively. \(G_{x,\text{filtered}}\) and \(G_{y,\text{filtered}}\) denote the gradients after filtering, with \(P_{Ux}\), \(P_{Lx}\), \(P_{Uy}\), and \(P_{Ly}\) specifying the upper and lower percentile thresholds for the X and Y directions. This approach allows for the emphasis on significant edge information while mitigating the impact of noise.

\subsection{Gaussian Blurring}
This process degrades the image with a Gaussian blur. This procedure specifies a kernel size of \(5 \times 5\) for blurring the image in the horizontal and vertical direction. And a standard deviation (\(\sigma\)) of 1 pixel is employed for both directions to maintain a consistent spread of the Gaussian distribution. The blurring equation is as follows:
\begin{align}
B &= (I_p * G_{\sigma})
\end{align}

where:
\begin{itemize}
    \item $B_x$ and $B_y$ are the images resulting from the convolution in the $x$ and $y$ directions, respectively.
    \item $I_p$ is the original image to which the blurring is applied.
    \item $G_{\sigma}$ represent the Gaussian kernel applied, with a standard deviation $\sigma$.
\end{itemize}

Following the blurring, we replicate the gradient computation and isolation process, but this time using the blurred versions of the original image, \(B_x\) and \(B_y\). Importantly, the same masks obtained from the original image, namely \(M_{LH}\), \(M_{P_x}\), and \(M_{P_y}\), are employed in this step. This ensures consistency in the criteria for selecting significant edges and reducing noise impact across both the original and blurred images. The gradients of the blurred images are calculated and filtered as:
\begin{align}
G_{B_x} &= \nabla_x(B_x \times M_{LH}) \times M_{P_x} \\
G_{B_y} &= \nabla_y(B_y \times M_{LH}) \times M_{P_y}
\end{align}
This approach not only involves computing the gradients \(G_{B_x}\) and \(G_{B_y}\) for the X and Y directions, respectively, after applying a Gaussian blur but also underscores the use of the same masks \(M_{LH}\), \(M_{P_x}\), and \(M_{P_y}\) obtained from the original image analysis. 

\subsection{Percentage Degradation Calculation}
In this analytical step, we engage in a comparative evaluation of the image's more pronounced gradients against its blurred version. This comparison involves the percentage difference between the filtered gradients, denoted as \(G_{x,\text{filtered}}\) and \(G_{y,\text{filtered}}\), and their Gaussian blurred equivalents, \(G_{B_x}\) and \(G_{B_y}\), for the \(x\) and \(y\) directions, respectively. Conceptually, the differences (\(G_{x,\text{filtered}} - G_{B_x}\) for the \(x\) direction, and similarly for \(y\)) illuminate the extent to which edge slope is compromised by the blurring effect. A notable degradation indicates a pronounced edge. This degradation is quantified by dividing the difference by the gradient's intensity, which yields the rate of edge decay in percentage terms for both the \(x\) and \(y\) directions:
\begin{align}
\Delta G_x &= \frac{G_{x,\text{filtered}} - G_{B_x}}{G_{x,\text{filtered}}} \\
\Delta G_y &= \frac{G_{y,\text{filtered}} - G_{B_y}}{G_{y,\text{filtered}}}
\end{align}
The division operation here serves a critical function: it renders the rate of decay independent of the edge's brightness and contrast. This effectuates a normalization of the decay rate that remains consistent across varying image contents. Consequently, \(\Delta G_x\) and \(\Delta G_y\) embody the percentage rate of edge decay, unaffected by the inherent brightness and contrast of the edge. This methodology affords a nuanced understanding of the impact that Gaussian blurring exerts on edge sharpness, offering insights into the edge's robustness or susceptibility to blurring influences, independent of the image's specific content features.

\subsection{Final Sharpness Metrics}
Finally, to obtain the sharpness values, we calculate the average of all differences between gradients. 
\begin{align}
S_x &= 100 \times \frac{1}{n} \sum_{i=1}^{n} \Delta G_{x_i}, \\
S_y &= 100 \times \frac{1}{n} \sum_{i=1}^{n} \Delta G_{y_i}.
\end{align}
This method involves computing the average decay of all gradients within the selected percentile ranges, leading to a single representative value. The final calculation involves multiplying the mean delta of the gradients by 100, as this result signifies a percentage. This percentage indicates the degradation level of the most pronounced edges within the image, providing a quantifiable measure of edge sharpness.

\subsection{Calculation of the Representativeness Indicator}
To  our analysis beyond sharpness, we introduce an indicator of representativeness that aims to quantify the extent to which an image meets the measurement criteria of the sharpness metric. This indicator seeks to quantify whether an image contains sufficient edges to be considered for sharpness measurement and whether its content is free from high-frequency elements or predominantly noise. In essence, the representativeness metric synthesizes into a singular numerical value the various scenarios in which an image might not accurately reflect the optical system, such as images obscured by clouds, lacking distinct edges, overly dark, or saturated images

\begin{figure}[htbp]
\centering
\begin{subfigure}[b]{0.31\linewidth}
    \includegraphics[width=\linewidth]{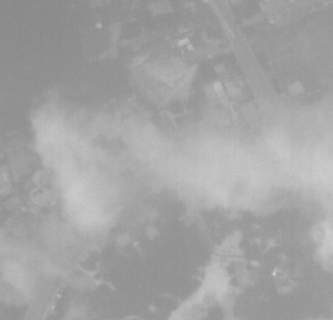}

    \label{fig:sigma_full_2}
\end{subfigure}
\hfill % Esto añade un espacio entre las figuras si es necesario
\begin{subfigure}[b]{0.31\linewidth}
    \includegraphics[width=\linewidth]{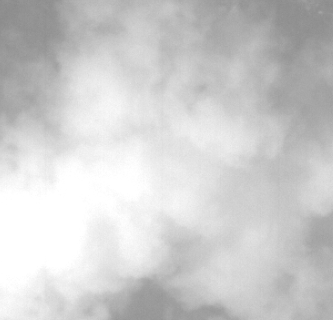}
   
    \label{fig:sigma_detail_1}
\end{subfigure}
\hfill
\begin{subfigure}[b]{0.31\linewidth}
    \includegraphics[width=\linewidth]{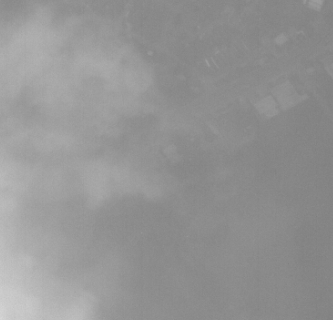}
    
    \label{fig:sigma_detail_2}
\end{subfigure}
\caption{Examples of images without enough edges}
\label{fig:benchmark_1}
\end{figure}

As first step, we avoid accounting for sensor noise and frequencies close to Nyquist, which generate excessively high sharpness values. We utilizes kernels scaled up threefold from the original Gaussian \(x\) and \(y\) kernels, coupled with a standard deviation (\(\sigma\)) of 5 pixels, aimed at enhancing the metric's sensitivity to the mentioned scenarios. The mathematical formulation of this step is as follows:
\begin{equation}
\text{BluredImage} = \text{GaussianBlur}\left(\text{Image}, \left(3 \cdot K_x, 3 \cdot K_y\right), 5\right)
\end{equation}

Following this, the representativeness indicator for the horizontal  and  vertical directions are calculated by first computing the gradients of the additionally blurred image, applying the low-high intensity mask (\(M_{LH}\)), and then filtering these gradients through percentile masks specific to each direction (\(M_{Px}\) and \(M_{Py}\)). The final metrics are obtained by averaging the values of these processed gradients, as expressed in the following equations, closely mirroring the approach utilized in sharpness analysis but adapted to evaluate representativeness:
\begin{equation}
\text{R}_x = \frac{1}{N} \sum \left( \nabla_x(\text{BluredImage} \times M_{LH}) \times M_{Px} \right)
\end{equation}
\begin{equation}
\text{R}_y = \frac{1}{N} \sum \left( \nabla_y(\text{BluredImage} \times M_{LH}) \times M_{Py} \right)
\end{equation}

Through these calculations, the representativeness metric offers a nuanced and quantifiable measure of an image's fidelity to the optical system it is intended to represent, accounting for potential deviations caused by various environmental and technical factors.

\subsection*{Sharpness Model Configuration:}
Consider $\mathcal{A}$ as a function that processes an input image $I$ with a set of configuration parameters $\mathcal{C}$, producing a set of metrics $\mathcal{M}$:

\[
\mathcal{M} = \mathcal{A}(I, \mathcal{C})
\]

where the set of configuration parameters $\mathcal{C}$ and the outcome metrics set $\mathcal{M}$ are defined as follows:

\[
\mathcal{C} = \{ C_{\text{percentiles}}, C_{\text{sobel}}, C_{\text{gauss}}, C_{\text{pixel\_dif}}, , C_{\text{low\_high\_threshold}} \}
\]

\[
\mathcal{M} = \{ M_{S_x}, M_{R_x}, M_{S_y}, M_{R_y} \}
\]

with each parameter in $\mathcal{C}$ specified as:

\begin{align*}
    &C_{\text{percentiles}}: \text{Upper and lower percentile thresholds for gradient analysis}, \\
    &C_{\text{sobel}}: \text{Sobel kernel size for gradient computation}, \\
    &C_{\text{gauss}}: \text{Gaussian kernel size and $\sigma$ for image smoothing}, \\
    &C_{\text{pixel\_dif}}: \text{Parameters for correcting anomalous pixels}, \\
    &C_{\text{low\_high\_threshold}}: \text{Parameters for masking low and high pixels}, \\
\end{align*}

The outcomes in $\mathcal{M}$, generated by the function $\mathcal{A}$, include:

\begin{align*}
    &M_{S_x}, M_{S_y}: \text{Sharpness measures in the X and Y directions, respectively}, \\
    &M_{R_x}, M_{R_y}: \text{Representativeness measures in the X and Y directions, respectively}.
\end{align*}

The parameters within $\mathcal{C}$ are subject to optimization based on the specific use case. The main factors to consider for the optimization of these parameters include the noise level in the images, the size of the images, the required confidence level of operation, among other relevant variables. The optimization process aims to balance these considerations to achieve the best possible performance of $\mathcal{A}$ in analyzing satellite imagery, ensuring that the derived metrics $\mathcal{M}$ accurately reflect the desired characteristics of sharpness and representativeness under varying conditions.

\subsection*{Sharpness Algorithm}

The pseudo code of the sharpness metric is the following:

\begin{algorithm}[H]
\caption{Enhanced Sharpness and Representativeness Analysis for Satellite Imagery}
\begin{algorithmic}[1]
\Require $Img , P_d, P_u, K_s, K_g, T_{LH}, P_{dif}$  \Comment{Input image and config parameters}
\Ensure $(S_x, R_x, S_y, R_y)$ \Comment{Sharpness and representativeness in X and Y}

\State $Img \gets \Call{FilterAnomalies}{Img, P_{dif}}$ \Comment{Filter anomalous pixels}
\State $Mask \gets \Call{CreateLowHighMask}{Img, T_{LH}}$ \Comment{Create mask for low/high pixels}
\State $G_x, G_y \gets \Call{ComputeGradients}{Img, Mask, K_s}$ \Comment{Compute gradients}
\State $SG_x, SG_y \gets \Call{SelectGradients}{G_x, G_y, P_d, P_u}$ \Comment{Select higher gradients}

\State $BImg \gets \Call{Blur}{Img, K_g}$ \Comment{Smooth the images}
\State $BG_x, BG_y \gets \Call{ComputeGradients}{BImg, Mask, K_s}$ \Comment{Gradients of smoothed images}

\State $\Delta G_x, \Delta G_y \gets \Call{ComputeBlurDiff}{SG_x, SG_y, BG_x, BG_y}$ \Comment{Compute blur differences}
\State $BImg_2 \gets \Call{Blur}{Img, K_g}$ \Comment{Create representative image}
\State $RG_x, RG_y \gets \Call{ComputeGradients}{BImg_2}$ \Comment{Gradients of representative image}

\State $R_x, R_y \gets \Call{ComputeMeans}{RG_x, RG_y}$ \Comment{Compute representativeness}
\State $S_x, S_y \gets \Call{ComputeMeans}{BD_x, BD_y} \times 100$ \Comment{Compute sharpness}

\Return $(S_x, R_x, S_y, R_y)$
\end{algorithmic}
\end{algorithm}

\section {Metric Validation}

The metric underwent evaluation through two distinct methodologies. The first method involved a high-volume, precision-driven approach using a dataset of simulated images. The second method employed real-world images of a significantly lower volume and without reference values.

\subsection{Simulated image benchmark}
The objective of the benchmark is to create numerous simulated images that are as realistic as those obtained from satellites. These images have variations in content, brightness, contrast, and noise, and are convolved with a real PSF from an actual satellite. Based on this, we convolve the image with a 9x9 Gaussian kernel with a variable sigma value, which will be the parameter correlated with the developed sharpness metric.
First, we generate 25000 realistic images with a resolution of $1000 \times 1000$ pixels. These images are designed to replicate the diversity in content, contrast, brightness, and noise levels characteristic of satellite acquired imagery. To achieve varied representations of content, the simulated images comprise squares measuring $n \times n$ pixels, with separations of $n$ pixels between them. This design strategy allows for the simulation of a broad spectrum of terrestrial features and textures, thereby enhancing the realism of the generated images. Additionally, we generated variations in contrast, brightness, and noise. The variations of the simulated images are as follows: Brightness, Contrast, Noise, Block Sizes and the blur introduced for prediction.

Subsequent to the generation of these content variations, the simulation introduces optical disturbances typical in satellite imaging. This is accomplished through the convolution of the images with a Point Spread Function (PSF) real kernel, simulating the minimal optical degradation inherent to satellite imaging systems. Furthermore, the simulation process incorporates different types of blur to mimic real-world imaging conditions. Specifically, a Gaussian blur is applied, employing a kernel size of $9 \times 9$. The simulation culminates with the addition of white Gaussian noise to the images, further augmenting their realism by simulating the random noise that affects actual satellite images due to factors such as atmospheric interference, sensor imperfections, and signal transmission disturbances. 

Crucially, the sigma values for these Gaussian kernels are correlated with the sharpness metric in the respective direction. This approach ensures a realistic simulation of how directional blur impacts image sharpness, independently of variations in contrast, brightness, and noise levels.

\begin{figure}[htbp]
\centering
\includegraphics[width=\linewidth]{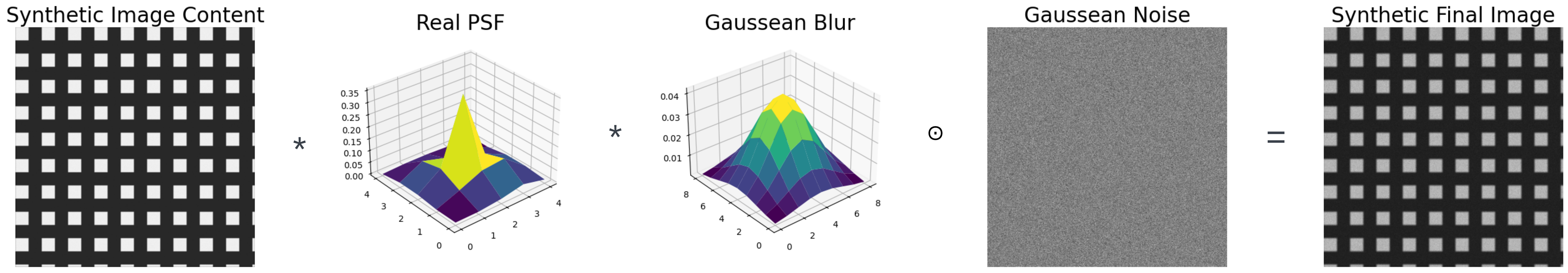}
\caption{Sythetic Generation Diagram} % Opcional
\label{fig:benchmark_2} 
\end{figure}

\begin{figure}[ht]
  \centering
  \begin{subfigure}{.5\textwidth}
    \centering
    \includegraphics[width=.9\linewidth]{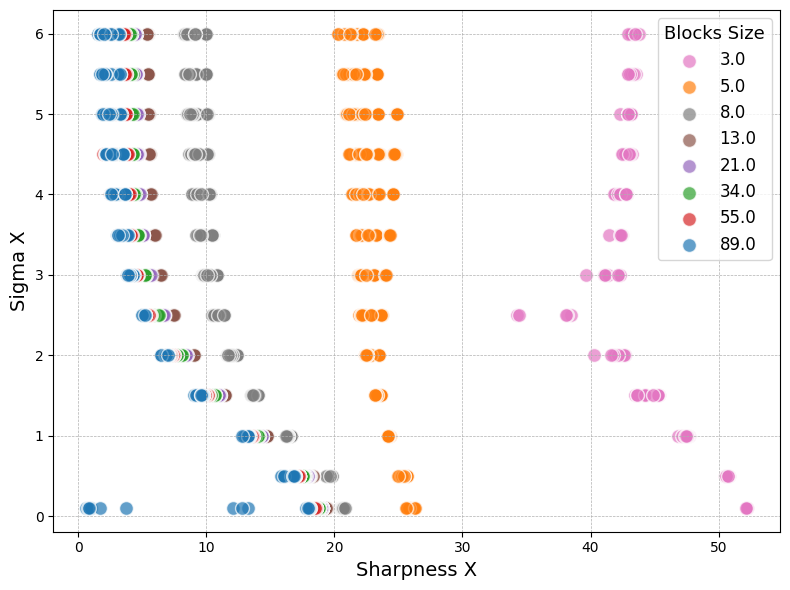}
    \caption{Unfiltered data}
    \label{fig:unfiltered}
  \end{subfigure}%
  \begin{subfigure}{.5\textwidth}
    \centering
    \includegraphics[width=.9\linewidth]{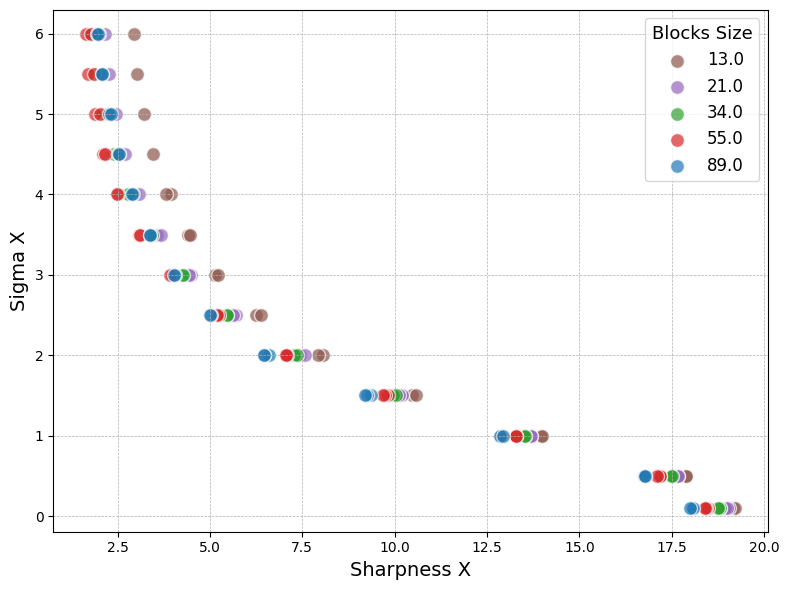}
    \caption{Filtered data using representativeness metric}
    \label{fig:filtered}
  \end{subfigure}
  \caption{Scatter plots of sharpness metric against $sigma$ levels. The size of each point corresponds to the block size, indicating its frequency content.}
  \label{fig:sharpness_plots}
\end{figure}

We present a scatter plot analysis (see Figure \ref{fig:sharpness_plots}) where the sharpness metric, denoted as $X$, is plotted against the $sigma$ levels of the image content.

The unfiltered data (Figure \ref{fig:unfiltered}) initially exhibited a high degree of variance with erratic behaviors, especially for smaller block sizes, implying an increased sensitivity to high-frequency content. Such observation suggests that the sharpness metric alone may not be sufficient for an accurate assessment under these conditions.

To address this, a representativeness metric was introduced, filtering out non-representative values. The resultant filtered data (Figure \ref{fig:filtered}) reveals a more discernible relationship between the sharpness metric and the different $sigma$ values. This segmentation facilitated by the representativeness metric underscores the metric's potential in distinguishing between varying levels of $sigma$ with improved precision.

\begin{figure}[htbp]
\centering
\begin{subfigure}[b]{0.31\linewidth}
    \includegraphics[width=\linewidth]{sigma001.png}
    \caption{Noise $\mu = 0.01$}
    \label{fig:sigma_full_1}
\end{subfigure}
\hfill % Esto añade un espacio entre las figuras si es necesario
\begin{subfigure}[b]{0.31\linewidth}
    \includegraphics[width=\linewidth]{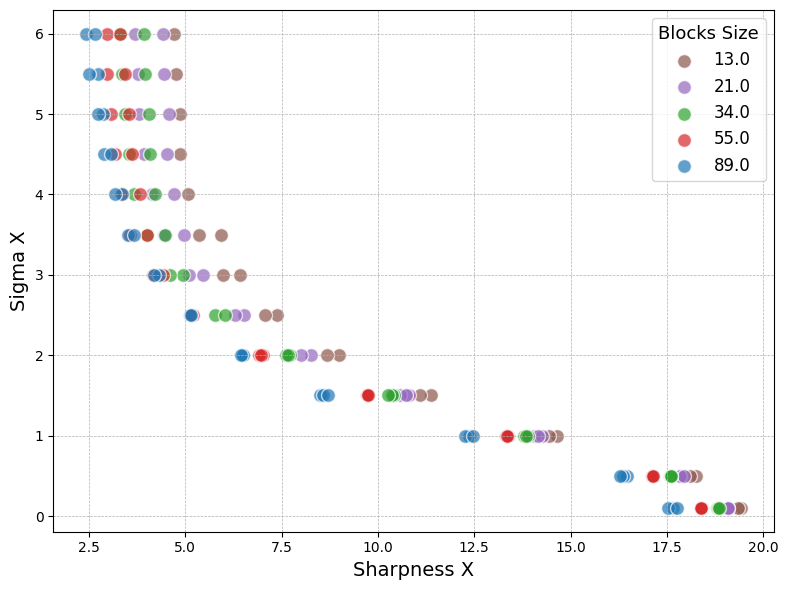}
    \caption{Noise $\mu = 0.03$}
    \label{fig:sigma_detail_3}
\end{subfigure}
\hfill
\begin{subfigure}[b]{0.31\linewidth}
    \includegraphics[width=\linewidth]{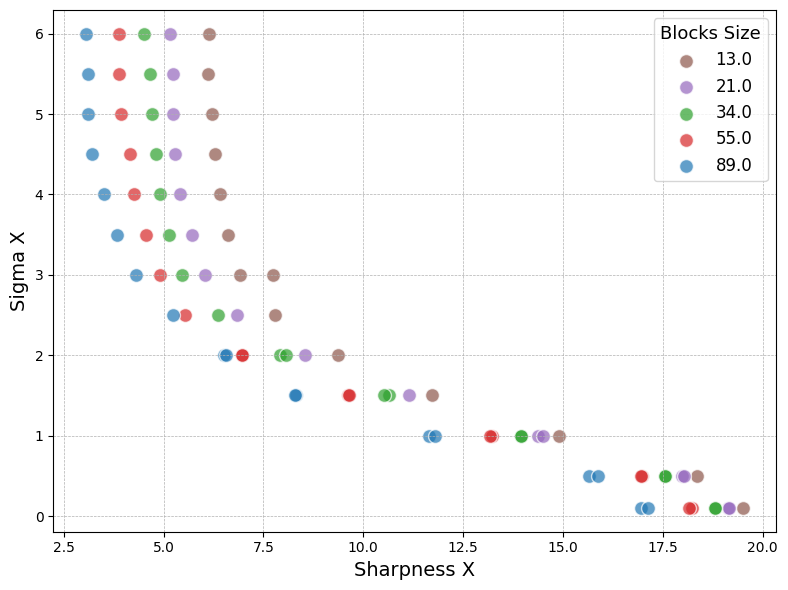}
    \caption{Noise $\mu = 0.05$}
    \label{fig:sigma_detail_4}
\end{subfigure}
\caption{Filtered Sharpness X metric vs Sigma in X direction, for different noise levels}
\label{fig:benchmark_3}
\end{figure}

The increase in standard deviation with higher sigma values reveals a compromise in the metric's ability to consistently represent image sharpness. The different variations in contrast and brightness do not appear to affect the estimation of sharpness, which is expected. Noise variations seem to increase the dispersion of the points, indicating an increase in uncertainty in the metric. The correlation between the metric and the sigma of the introduced blur is sufficiently good. We can also observe that the metric shows greater sensitivity to images with higher sharpness compared to images with lower sharpness.

\subsection{Real image benchmark}
To test the metric with real images, we conducted two experiments where we captured several photos from the same vantage point, varying the focus position so that the images transitioned from being out of focus to in focus and then back to being out of focus. The experiments were performed using parameters tuned specifically for our images. As an example, we present two sequences of photos with different content. The focus position changes over time, and we expect the metric to accurately reflect these changes in focus position as a curve. Multiple images were captured for each focus position. The results are as follows:

\begin{figure}[H]
    \centering
    \begin{minipage}{0.45\textwidth}
        \centering
        \includegraphics[width=\linewidth]{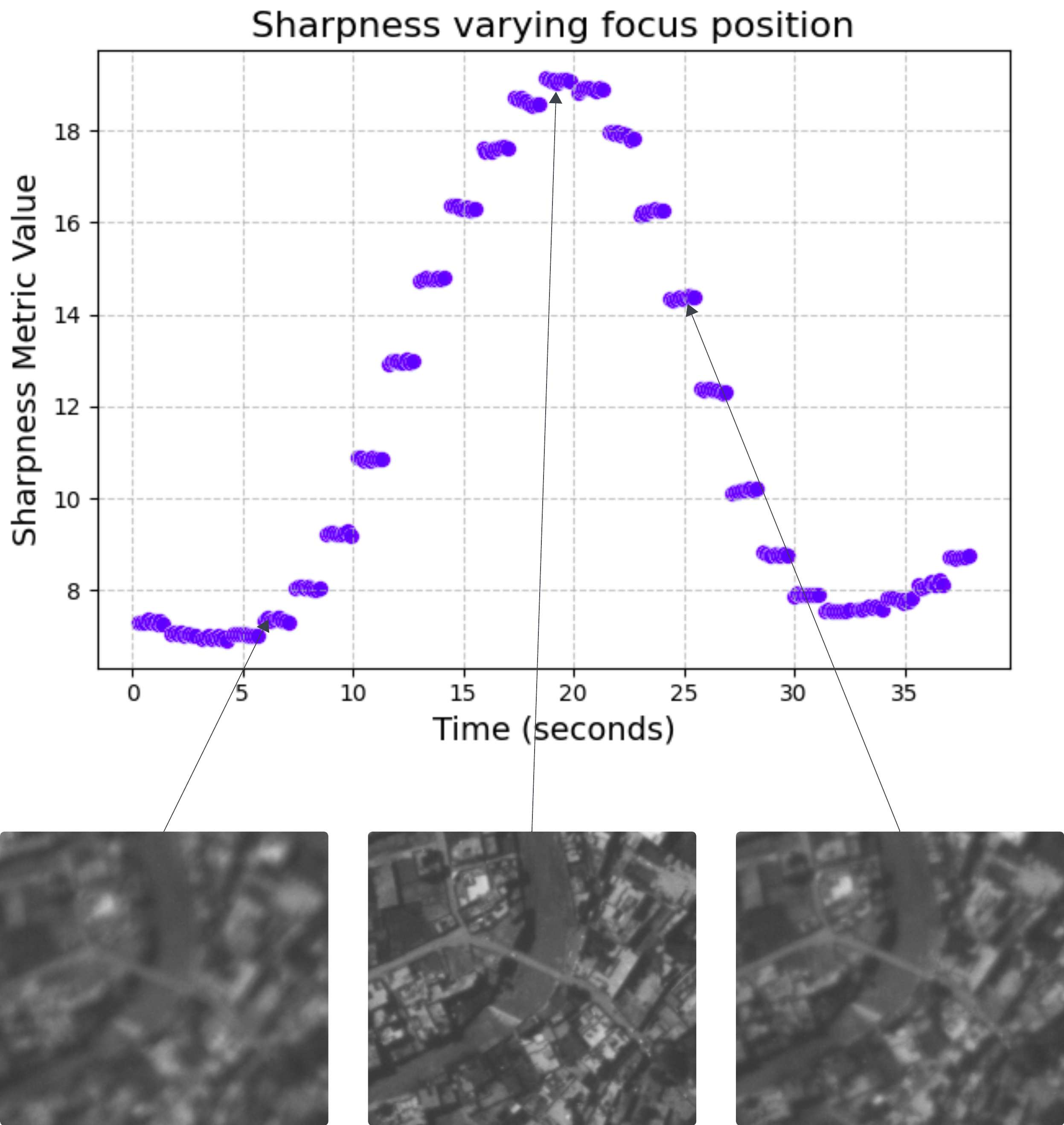}
        \caption{Results of experiment 1}
        \label{fig:experiment_1}
    \end{minipage}
    \hfill
    \begin{minipage}{0.45\textwidth}
        \centering
        \includegraphics[width=\linewidth]{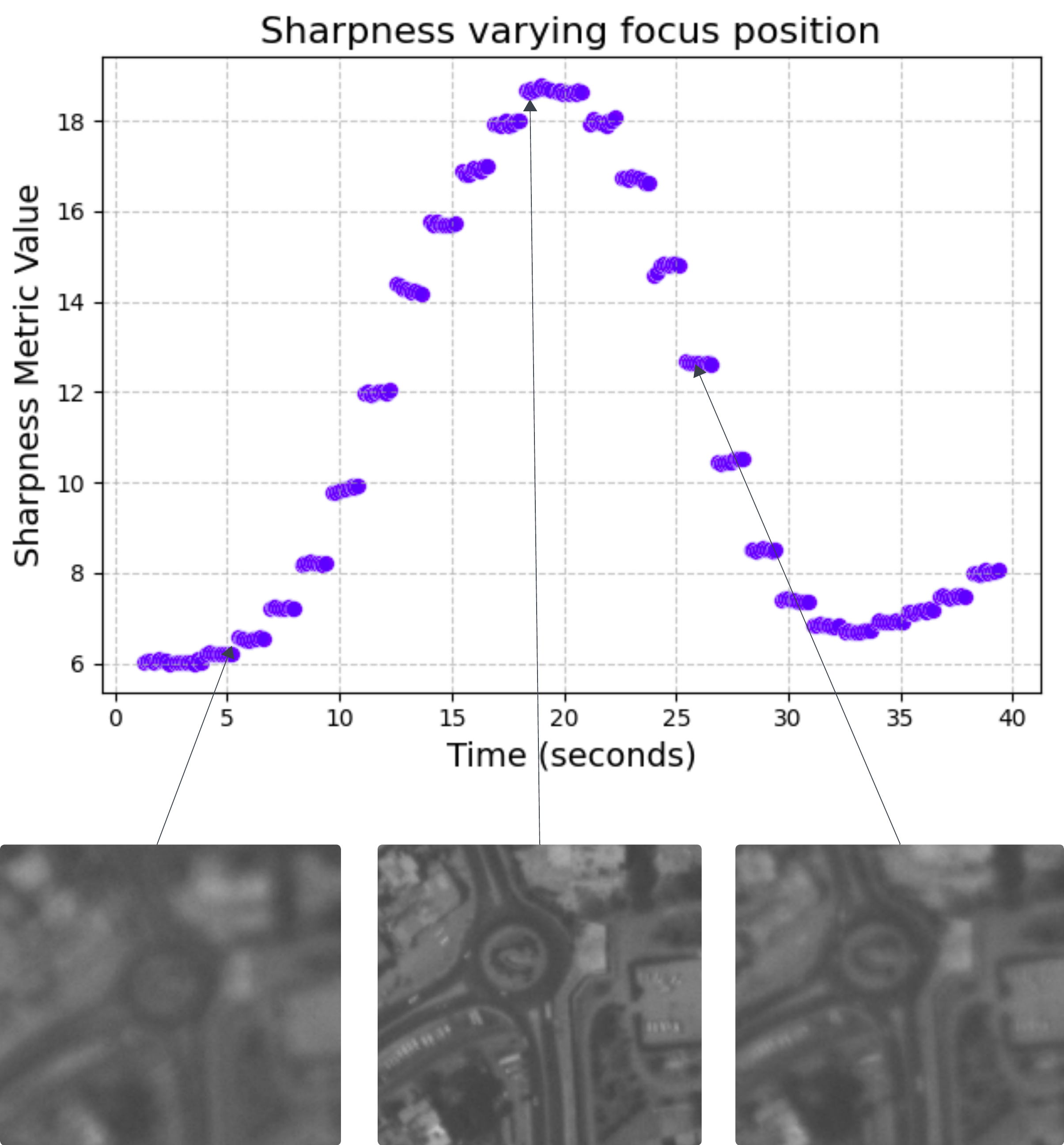}
        \caption{Results of experiment 2}
        \label{fig:experiment_2}
    \end{minipage}
\end{figure}

In the graphs, we can observe that if the focus position is not varied, the metric score for each image remains relatively constant. However, there are noticeable steps indicating where the focus position likely changes. Additionally, the maximum quality points are very close in both experiments, with 19.1 for Experiment 1 and 18.9 for Experiment 2. It is also evident that the sharpness of the observed images correlates correctly with the metric values.

\section*{Conclusion}

In this study, we have introduced a content-robust metric designed specifically for assessing the quality of satellite imagery. This innovative metric accurately quantifies the sharpness levels in the X and Y directions of the acquisition system, providing a critical tool for the analysis and enhancement of image quality obtained from the satellites.

The significance of our proposal lies in its remarkable resistance to variations in brightness, contrast, and image content, in addition to its robustness against various types of noise. This attribute is crucial, as it ensures the reliability and precision of the metric across a wide spectrum of conditions, thereby facilitating an objective and consistent evaluation of satellite image quality.

The implementation of our metric represents a significant advancement in the ability to monitor and quantify, in an automated manner, the quality at which different satellites operate. This development not only optimizes the performance of image acquisition systems but also promotes the standardization of satellite image quality.

In conclusion, the metric formulated within the scope of this investigation represents a progressive advancement in the pursuit of precise and automated methodologies for the evaluation of satellite image quality. The ability of this metric to function effectively under a diverse array of conditions underscores its critical importance as an essential instrument within the technological repertoire for Earth observation. This, in turn, fosters substantial enhancements in the acquisition and analysis of geospatial data.


\begin{thebibliography}{99}
\bibitem{Brenner1976} Brenner, J. F., et al. (1976). An automated microscope for cytologic research: A preliminary evaluation. \textit{The Journal of Histochemistry and Cytochemistry}.
\bibitem{Marziliano2002} Marziliano, P., et al. (2002). A no-reference perceptual blur metric. In \textit{Proceedings of the IEEE International Conference on Image Processing}.
\bibitem{Zhou2018} Zhou, W., et al. (2018). Blind image quality assessment based on learning to rank. \textit{IEEE Transactions on Image Processing}.
\bibitem{Li2016} Li, X., et al. (2016). A multiple level set method for color image segmentation using a modified GAC model. \textit{Computers Electrical Engineering}.
\bibitem{Zhan2020} Zhan, Y., et al. (2020). No-reference image sharpness assessment based on maximum gradient and variability of gradients. \textit{IEEE Transactions on Multimedia}.
\bibitem{Li2021} Li, Y., et al. (2021). Blind image quality assessment using statistical structural and luminance regularity. \textit{IEEE Transactions on Multimedia}.
\bibitem{Canny1986} Canny, J. (1986). A computational approach to edge detection. \textit{IEEE Transactions on Pattern Analysis and Machine Intelligence}.
\bibitem{Sobel1970} Sobel, I. (1970). Camera Models and Machine Perception. \textit{Stanford Artificial Intelligence Project}, Technical Report.
\bibitem{Prewitt1960} Prewitt, J. M. S. (1970). Object enhancement and extraction. In \textit{Picture Processing and Psychopictorics}, Academic Press.
\bibitem{Zhang2016} Zhang, X., et al. (2016). Edge-preserving image smoothing with adaptive clustering. \textit{IEEE Transactions on Image Processing}.
\bibitem{Kanjar2016} Kanjar, D., et al. (2016). Image sharpness measure for blurred images in frequency domain. \textit{Procedia Engineering}.
\bibitem{Bae2018} Bae, S. H., et al. (2018). A novel image quality assessment with globally and locally consilient visual quality perception. \textit{IEEE Transactions on Image Processing}.
\bibitem{Baig2022} Baig, M. A., et al. (2022). DFT-based no-reference quality assessment of blurred images. \textit{Multimedia Tools and Applications}.
\bibitem{Kerouh2018} Kerouh, F. (2018). A no reference quality metric for measuring image blur in wavelet domain. \textit{International Journal of Digital Information and Wireless Communications}.
\bibitem{Vu2011} Vu, C. T., et al. (2011). S3: A spectral and spatial measure of local perceived sharpness in natural image. \textit{IEEE Transactions on Image Processing}.
\bibitem{Hassen2013} Hassen, R., et al. (2013). Image sharpness assessment based on local phase coherence. \textit{IEEE Transactions on Image Processing}.
\bibitem{Wang2018} Wang, Z., et al. (2018). A novel image sharpness assessment based on spatial frequency and multiscale gradient. \textit{IEEE Transactions on Multimedia}.
\bibitem{Gvozden2018} Gvozden, G., et al. (2018). Blind image sharpness assessment based on local contrast map statistics. \textit{Journal of Visual Communication and Image Representation}.
\bibitem{Pei2015} Pei, B., et al. (2015). A no-reference image sharpness metric based on large-scale structure. \textit{Journal of Physics: Conference Series}.
\bibitem{Liu2016} Liu, X., et al. (2016). No-reference image quality assessment with shearlet transform and deep neural networks. \textit{Neurocomputing}.
\bibitem{Zhu2018} Zhu, H., et al. (2018). Image quality assessment based on deep learning with FPGA implementation. \textit{Signal Processing: Image Communication}.
\bibitem{Lin2020} Lin, L., et al. (2020). A novel scheme for image sharpness using inflection points. \textit{International Journal of Imaging Systems and Technology}.
\bibitem{Moorthy2011} Moorthy, A. K., et al. (2011). A two-step framework for constructing blind image quality indices. \textit{IEEE Signal Processing Letters}.
\bibitem{Li2020} Li, D., et al. (2020). Exploiting high-level semantics for no-reference image quality assessment of realistic blur images. In \textit{Proceedings of the ACM International Conference on Multimedia}.
\bibitem{Gao2019} Gao, F., et al. (2019). Blind image quality prediction by exploiting multi-level deep representations. \textit{Pattern Recognition}.
\bibitem{Xu2018} Xu, J., et al. (2018). Blind image quality assessment based on high order statistics aggregation. \textit{IEEE Transactions on Image Processing}.
\bibitem{Jiang2019} Jiang, Q., et al. (2019). Optimizing multistage discriminative dictionaries for blind image quality assessment. \textit{IEEE Transactions on Multimedia}.
\bibitem{Liu2018} Liu, S., et al. (2018). Salient region guided blind image sharpness assessment. \textit{Sensors}.
\bibitem{Zhu2023} Zhu, M. L., et al. (2023). Image quality assessment based on deep learning with FPGA implementation. \textit{Signal Processing: Image Communication}.
\bibitem{Mittal2012} Mittal, A., et al. (2012). No-reference image quality assessment in the spatial domain. \textit{IEEE Transactions on Image Processing}.
\end{thebibliography}
\end{document}